# Experimental Investigation of Surface Passivation Chemistries for Optical Nanotweezers

Maxwell T. Ugwu*†, Kewei. Li†#, Abayomi. Opadele†#, Theodore Anyika†#, and Justus C. Ndukaife†*#

*Department of Interdisciplinary Material Science, Vanderbilt University, Nashville TN, USA 37215

† Vanderbilt Institute for Nanoscale Science and Engineering, Vanderbilt University, Nashville TN, USA 37235

#Department of Electrical and Computer Engineering, Vanderbilt University, Nashville TN, USA 37215

**ABSTRACT:** Nanotweezers are actively investigated as a powerful means to reversibly trap and characterize nanoparticles with profound biological and environmental importance. To ensure that these particles can be reversibly trapped and released, surface passivation is essential. To mitigate the issue of fouling, we investigated the antifouling properties of poly(sodium styrene sulphate) (PSS) synthesized using the Atom Transfer Radical Polymerization (ATRP) technique on our previously reported gold-based Interferometric Electrohydrodynamic Tweezers (IET) device. Fluorescence and interferometric scattering (ISCAT) imaging were used to record trapping performance and study the antifouling properties of our passivated nanotweezer device. The results show that PSS exhibits superior anti-fouling performance against polystyrene nanoparticles when compared to 11-mercaptoundecanoic acid (MUA). By comparing the antifouling properties of PSS and zwitterionic poly(methacryloyloxyethyl phosphorylcholine) (PMPC) in preventing extracellular vesicle adhesion, we found that both exhibited similar performance. Overall, the ATRP technique is broadly applicable across nanotweezer substrates with appropriately chosen initiators.

In recent years, nanotweezers have emerged as a transformative platform for the precise trapping, manipulation, and sorting of nanoscale objects [1-6]. This technique has enabled unprecedented opportunities to investigate nanoscale biological and synthetic entities that are far beyond the ability of traditional diffracted-limited laser tweezers. Recently, gold-based nanoaperture optical tweezers have demonstrated remarkable performance in the high-precision classification of cancerous extracellular vesicles, showcasing their strong potential for biomedical diagnostics and single-particle analysis [7].

Gold and silicon are extensively used in the fabrication of nanotweezers because they support strong highly confined plasmonic and Mie resonances, respectively. These resonances enable the confinement of light to the nanoscale for efficient nano-optical trapping. While dielectric materials like silicon have been employed for optical trapping of nanoparticles due to their relatively low intrinsic optical losses compared to metals, gold remains the most widely studied material for most nanotweezer devices in the literature. This preference is driven by gold's outstanding resistance to environmental degradation. The intrinsic heat generation associated with plasmonic nanotweezers (such as those made with gold), and the associated temperature rise can be managed by the integration of a heat sink comprising a high thermal conductivity substrate like crystalline silicon [8]. When rapid transport of target particles is desired, electrothermoplasmonic flow generated by laser-induced thermal gradients can be harnessed for fast transport [9].

In addition to particle manipulation, plasmonic nanotweezers are enabling new possibilities in label-free single molecule sensing. For example, a gold nanoaperture optical tweezer has enabled direct mapping of the conformational energy landscape of a single unmodified protein [10]. Lately, our group reported an innovative approach that enables the simultaneous trapping of numerous particles in parallel facilitated by A.C. electro-osmotic (ACEO) flow. These platforms are known as geometry-induced electrohydrodynamic tweezer (GET) [11] and interferometric electrohydrodynamic tweezer [12], which leverage A.C. electro-osmotic (ACEO) flow to achieve massively parallel trapping of nanoparticles and biological objects. These platforms represent a substantial advancement toward scalable, high-throughput nanoscale manipulation technologies.

To enable robust operation of nanotweezer-based ultrafast nanoparticle trapping and characterization, it is imperative to mitigate non-specific adsorption. Non-specific adsorption of nanoparticles onto device surfaces often leads to irreversible particle adhesion, and inability to reuse chips after trapping experiments. To effectively operate nanotweezer devices, it is imperative to prevent non-specific adsorption of the target particles to the surface of the nanotweezer devices. Typically, researchers have mostly added surfactants to the trapping solution to minimize non-specific adsorption [13]. However, this approach remains fundamentally limited because it does not directly address the surface chemistry of the nanotweezer device itself and may adversely impact particles like biological vesicles. In this work, we focus on investigating molecular layers to passivate the nanotweezer surfaces and tune the interplay between Van der Waals attraction and electrostatic repulsion.

Molecules with thiol(-SH) functional group have been reported for the surface passivation of gold-based biosensor devices due to their strong affinity towards Au surfaces [14, 15]. With their thiol head group, these molecules bond stably to Au via physisorption while preventing fouling with their terminal group (Figure 1b). Fouling reduction is accomplished through the effect of steric hindrance and the ability of the terminal group to form a hydration layer [16] and/or have a net electric charge. The charge and hydrophilicity of the terminal groups determines the integrity of the hydration layer and repulsive force, while the steric hindrance is attributed to the stable configuration of organic carbon chain. The hydration layer formed by these molecules and their structural conformation has been proven to be influenced by microenvironments characterized by interfacial pH and surface roughness [17]. In a broader sense, antifouling molecules are classified based on their terminal group, and a variety of these molecules have been explored extensively for different applications.

Molecules with ethylene glycol terminal groups poly(ethylene glycol) (PEG) [18,19], and oligo (ethylene glycol) (OEG) [20] have been the most popular choice for fouling reduction for different variety of surfaces and applications. Several passivation strategies have been investigated, including polymer brush formation and high-packing-density self-assembled monolayers (OEG-SAMs) [21–24], which provide a tightly bound hydration layer [25]. However, the ethylene glycol groups in these systems are inherently unstable and prone to oxidative degradation [25]. This limitation has motivated increasing interest in the more stable zwitterionic compounds, which can bind water molecules even more strongly via electrostatically induced hydration [25, 26]. Other hydrophilic materials such as dextran [27] mannitol [28], Hyaluronic Acid [29] have also been explored. Although these materials have demonstrated promising antifouling characteristics in biosensing applications, their effectiveness and compatibility with nanotweezer devices, particularly in the fouling reduction of polymer nanoparticles and extracellular vesicles, which have garnered significant attention in recent years—remains largely unknown.

In this work, we employed the Atom Transfer Radical Polymerization (ATRP) technique, to grow poly(sodium styrene sulphate) (PSS) layers directly on our interferometric electrohydrodynamic tweezer (IET) device. The antifouling performance of the ATRP grown PSS is evaluated and compared against the treatment with 11-Mercaptoundecanoic acid for fluorescence polystyrene, and the surface coated with poly(methacryloyloxyethyl phosphorylcholine) (PMPC) for extracellular vesicles. While the present study primarily centers on IET as the nanotweezer platform, these findings are transferable and can be applied to any gold-based nanotweezer device. With the appropriate ATRP initiators this methodology can also be extended to a wide range of nanotweezer substrates such as silane initiators for silicon and silicon dioxide, and phosphate initiators for indium tin oxide, thereby providing a versatile pathway to universal anifouling nanotweezer interfaces.

### Working principle of interferometric electrohydrodynamic tweezers (IET)

The IET functionalized with an antifouling layer such as 11-Mercaptoundecanoic acid (MUA) is illustrated in Figure 1a. The device features a gold film with a 15 μm hole array having an edge-to-edge spacing of 2 μm between adjacent holes. An Indium Tin Oxide (ITO) electrode is placed over the patterned gold film with a dielectric spacer layer forming a microfluidic channel. An AC voltage is applied across the patterned gold film. The applied voltage generates normal and tangential electric fields. The tangential field interacts with the diffuse electric double layer charge at the solid-fluid interface, driving surface ACEO flows which localize single nanoscale objects in the interstitial sites of the 15 μm hole array (Figure 1a). The velocity of the ACEO flow is given by the Helmholtz-Smoluchowski slip velocity. [30]:

$$u_s = -\frac{\varepsilon_w \zeta}{\eta} E_{\parallel}$$

where $u_s$ is the velocity of the ACEO flow, $\varepsilon_w$ is the permittivity of the fluid medium, $\zeta$ is the zeta potential, $\eta$ is the fluid viscosity, and $E_{\parallel}$ is the tangential component of the electric field.

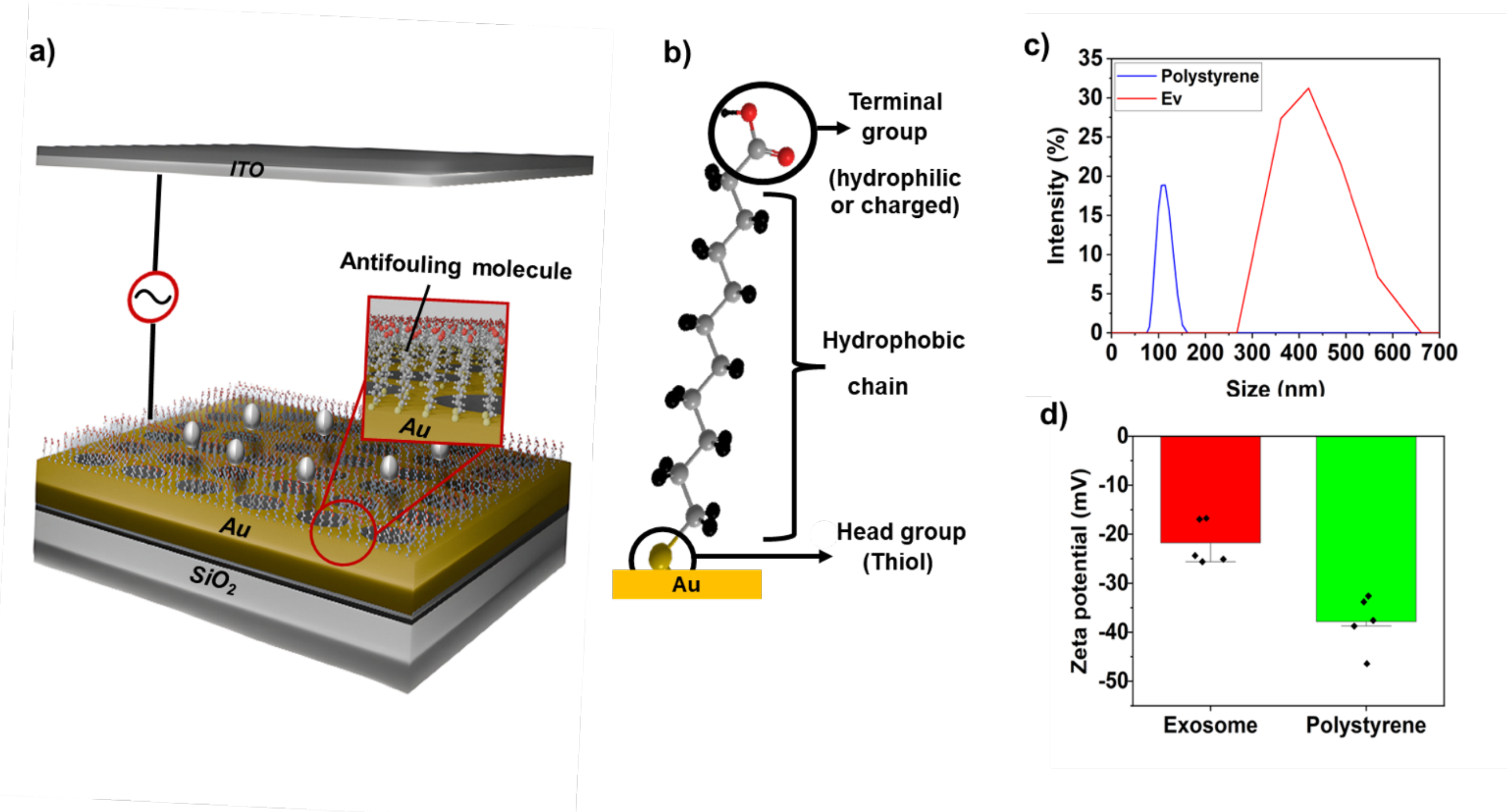

**Figure 1: a)** The IET functionalized with 11-Mercaptoundecanoic acid (MUA), **b)** Labeled chemical structure of MUA on a gold film, **c)** Particle size distribution and **d)** Zeta potential of the polystyrene polymer beads and extracellular vesicle (EVs) sample used for the antifouling study.

## Antifouling studies and surface characterization

We first begin by trapping polystyrene polymer beads with an unpassivated nanotweezer chip. A high concentration ($10^{10}$ particle/ml) of fluorescently labeled polymer beads with a diameter of 100 nm is introduced into the unpassivated chip. The trapping was achieved under an applied AC frequency of 3 kHz and amplitude of 10 V. In Figure 2a, some of the trapped particles are highlighted with green circles (left frame) and the adsorbed particles are highlighted with red circles (both left and right frames). From the result in Figure 2a, it is evident that within only 27 minutes of trapping experiments, particles spontaneously adhere to the chip's surface due to the Van der Waals force of attraction between the surface and the polystyrene particles for the unpassivated IET device. Within this time frame, a significant number of particles have adhered to the surface, adversely affecting the concentration of particles recovered following characterization or sorting experiments.

To mitigate this issue of fouling for nanotweezers, several factors must be considered such as the media pH and composition, the nanoparticle's zeta potential (Figure 1d), the appropriate molecule with the right terminal group for antifouling, hydrophobic chain for high parking density and the head group for stable bonding to the substrate. The anti-fouling 11-Mercaptoundecanoic acid (MUA) can hinder access of redox ions to the gold surface [31], owing to its capability of stable bonding to gold with the thiol head group and ability to form a self-assembled monolayer (SAM) with the hydrophobic carbon chain. Additionally, at pH 7 and above, the hydrophilic (COOH) terminal group in MUA can deprotonate to form a net negative charge [32], which can be useful in repelling negatively charged particles. Moreover, the (COOH) terminal group enables MUA to serve as a linkage molecule between Au surface and biorecognition elements, such as aptamers [33] or antibodies [34].

Considering the factors above, we passivated the IET surface with 11-Mercaptoundecanoic acid (MUA), and at a media of pH 7 (deionized water) we investigated the rate of adhesion of 100 nm fluorescence labeled polystyrene beads with a zeta potential of -40 mV**.** The MUA antifouling layer performance data in reducing polystyrene adsorption after 40 minutes is shown in Figure 2b. From the figure, the fouling of polystyrene was reduced significantly and only four particles adhered to the surface within the working field of view. This result is attributed to the high packing density of the self-assembled monolayer formed by the hydrophobic alkyl chains, the hydrophilic nature of the carboxyl group, and most importantly the repulsive force induced on negatively charged polystyrene (Figure 1d) by the deprotonated carboxyl group at pH ~7. This result is significant as it opens the door to efficient trapping and release of nanoparticles such as nanoplastics and other negatively charged particles with minimal fouling, thereby enabling their recovery following sorting and/or characterization.

We also investigated the antifouling performance of ethylene glycol molecules namely, 1-Mercaptoundec-11-yl) tetra (ethylene glycol) (OEG) and 1-Mercapto-11-hydroxy-3,6,9-trioxaundecane (TMOL) in the Supplementary section. These materials showed minimal reduction in polystyrene bead fouling because they lack surface electric charge (Supplementary movies 5 and 6).

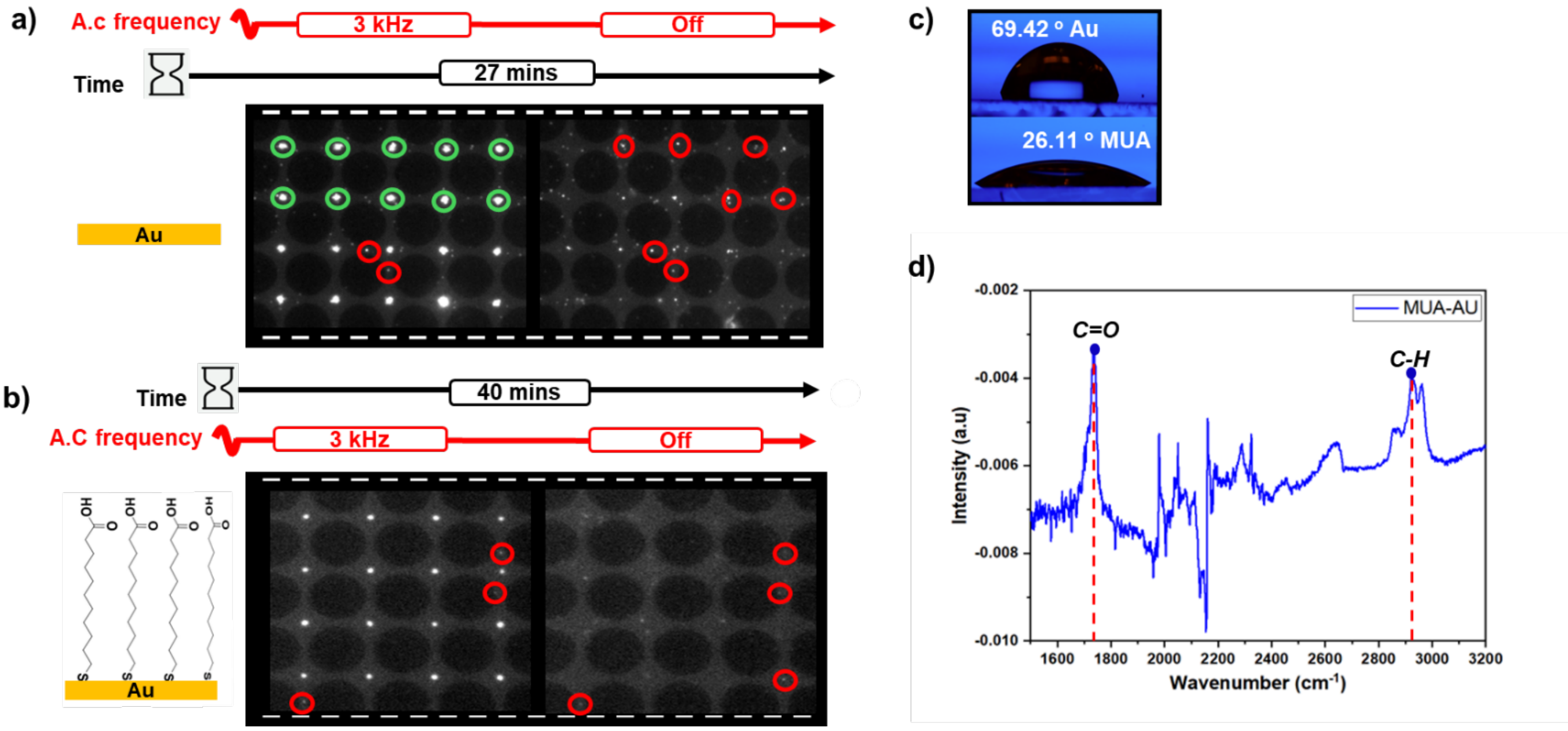

**Figure 2:** Fluorescence images of trapped and released polystyrene with **a)** unpassivated IET and **b)** IET passivated with 11-Mercaptoundecanoic acid (MUA). The green circles highlights some of the trapped particles at 3 kHz while the red circles highlight some adsorbed particles due to fouling **c)** Contact angle result of the bare and MUA functionalized IET and **d)** Infrared spectrum of 11-Mercaptoundecanoic acid (MUA) on gold.

The hydrophilicity and infrared vibrational properties of the bare and MUA passivated nanotweezer chip surface was investigated using contact angle analysis and ATR-FTIR spectroscopy (Figure 2c-d). From Figure 2c, the contact angle of water on the IET chip before passivation shows a high value of 69.42 degrees owing to the relative hydrophobic nature of the surface. After passivation with MUA, the value dropped by more than a factor of two due to the presence of the hydrophilic terminal group (COOH) [35]. The infrared spectrum showed the presence of acid C=O stretching and carbon chain C-H stretching band at 1740 $cm^{1}$ and 2850 $cm^{-1}$ respectively [36], confirming the presence of MUA on the IET surface.

Although MUA improved the antifouling performance compared to unpassivated IET (Figure 2b) due to its negative charge, its effectiveness is limited to a narrow pH range [32], and it provides only a single layer of deprotonated ($COO^{-}$) groups for particle repulsion. To further improve the antifouling properties of the IET nanotweezer device, we used the classical Atom Transfer Radical Polymerization (ATRP) reaction method as discribed in [25] to grow multilayer hydrophilic and negatively charged poly(sodium styrene sulphate) (PSS) on a IET nanotweezer chip with 11-Mercaptoundecyl 2-Bromo-2-methylpropanoate initiator. With the initiator's thiol head group stably bonded to gold, this enables the solid attachment of the introduced antifouling film to ensure the reusability of the chip. This technique, as shown in Figure 3, uses Cu(I)Br catalysts and ligand (2,2 bipyridyl) under nitrogen protection to activate the polymer chains/initiator, allowing for the polymerization of materials with multiple-layer film on a self-assembled monolayer initiator for efficient packing. PSS and PMPC have been successfully synthesized on the IET surface with ATRP, and their FTIR spectrum are presented in Figure 3b and 3c.

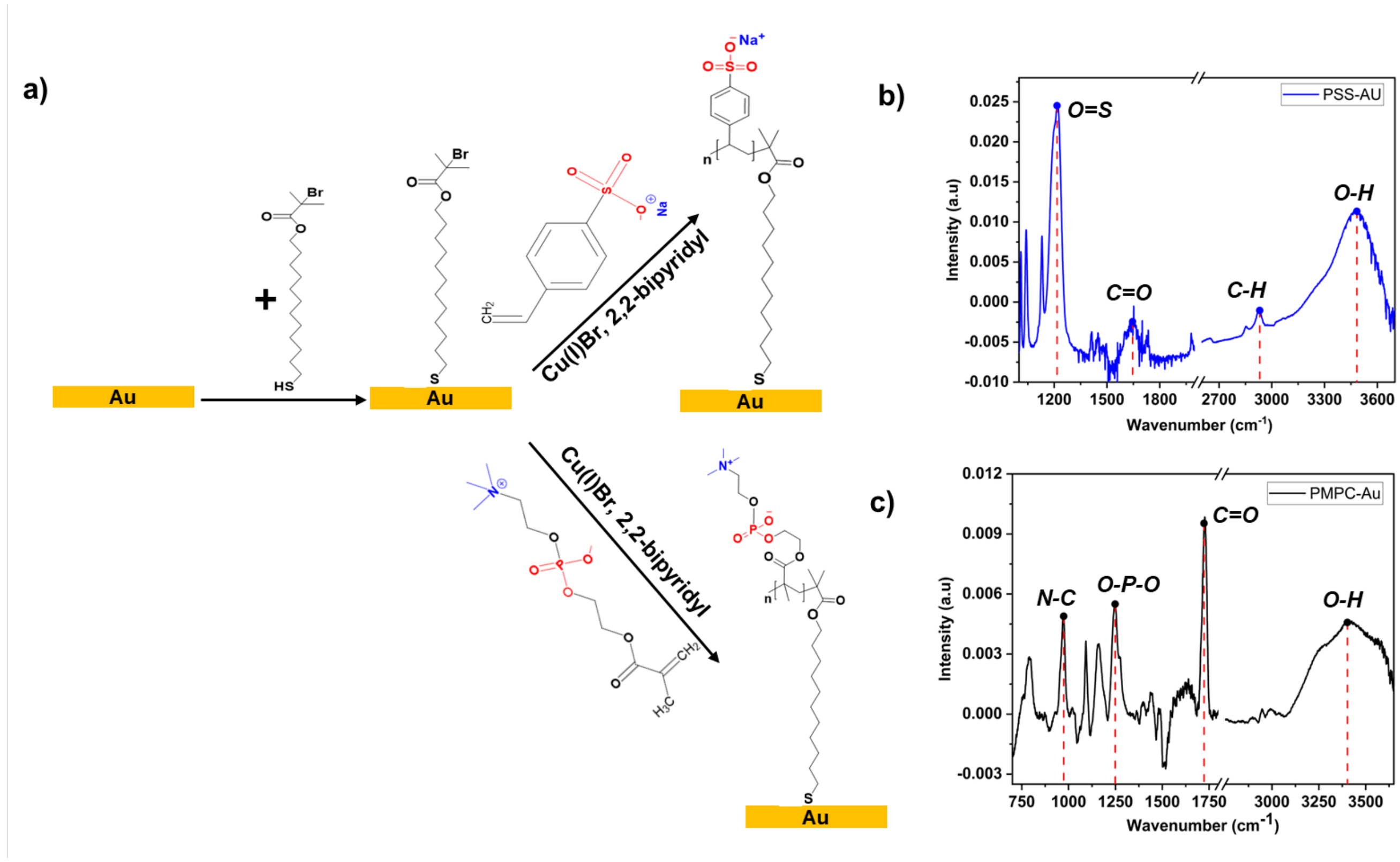

Figure 3: Classical Atom Transfer Radical Polymerization (ATRP) reaction steps for synthesizing **a)** poly(sodium styrene sulphate)(PSS) and poly(methacryloyloxyethyl phosphorylcholine) (PMPC) on gold surfaces. **b)** IR spectrum of PSS and **c)** PMPC.

The passivation of our IET device with terminal groups tailored to reduce fouling of different classes of particles demonstrates the versatility of ATRP-based surface passivation, as shown in Figure 4. Using the same functionalization strategy and initiator (same substrate), while varying the monomer, we employed PSS to reduce fouling of polystyrene and PMPC to minimize extracellular vesicle fouling. Finally, we compared the antifouling performance of PMPC and PSS against extracellular vesicles. It is important to note that molecules containing thiol head groups together with these specific terminal functionalities are generally not commercially available. Consequently, ATRP-based surface functionalization provides significant flexibility by enabling independent selection of the head group appropriate for a given substrate and the terminal group tailored to the requirements of a specific application or project.

Figure 4a presents trapping and releasing results (from left to right frame) for 100 nm polystyrene in an IET passivated with poly(sodium styrene sulphate) (PSS). As shown in Figure 4a, only a single polystyrene particle remained adhered to the surface after 40 minutes of experimentation. Notably, this particle was attached to an exposed region of the substrate rather than to the passivated gold film, while the remaining particles visible in the figure were freely diffusing away from the trap (Supplementary Movie 4). This shows improved antifouling property over MUA due to the multiple negatively charged styrene sulfonate layers formed using the ATRP process.

For extracellular vesicle antifouling studies, we employed the interferometric scattering imaging technique (Supplementary Figure 1) to measure the particle adsorption rate on our passivated and unpassivated nanotweezer device. The interferometric imaging enables label-free imaging since the EVs are unlabeled. Figure 4c (Supplementary Movie 1) presents interferometric scattering images of trapped and released (from left to right frame) EVs using the unpassivated IET. From this Figure, after 40 minutes, most of the trapped particles were adhered on the gold film after release as highlighted with the red circles. Figures 4b (Supplementary Movie 3) and 4d (Supplementary Movie 2) present the antifouling results, highlighting the performance of negatively charged PSS and neutral zwitterionic PMPC in reducing EV fouling. The number of particles in the released frame is significantly lower than that of unpassivated counterpart. Since EVs exhibit a negative zeta potential as shown in Figure 1c, PSS gave similar antifouling result when compared to PMPC due to the hydrophilic nature and the repulsive force induced by the negatively charged multilayer sulphate groups.

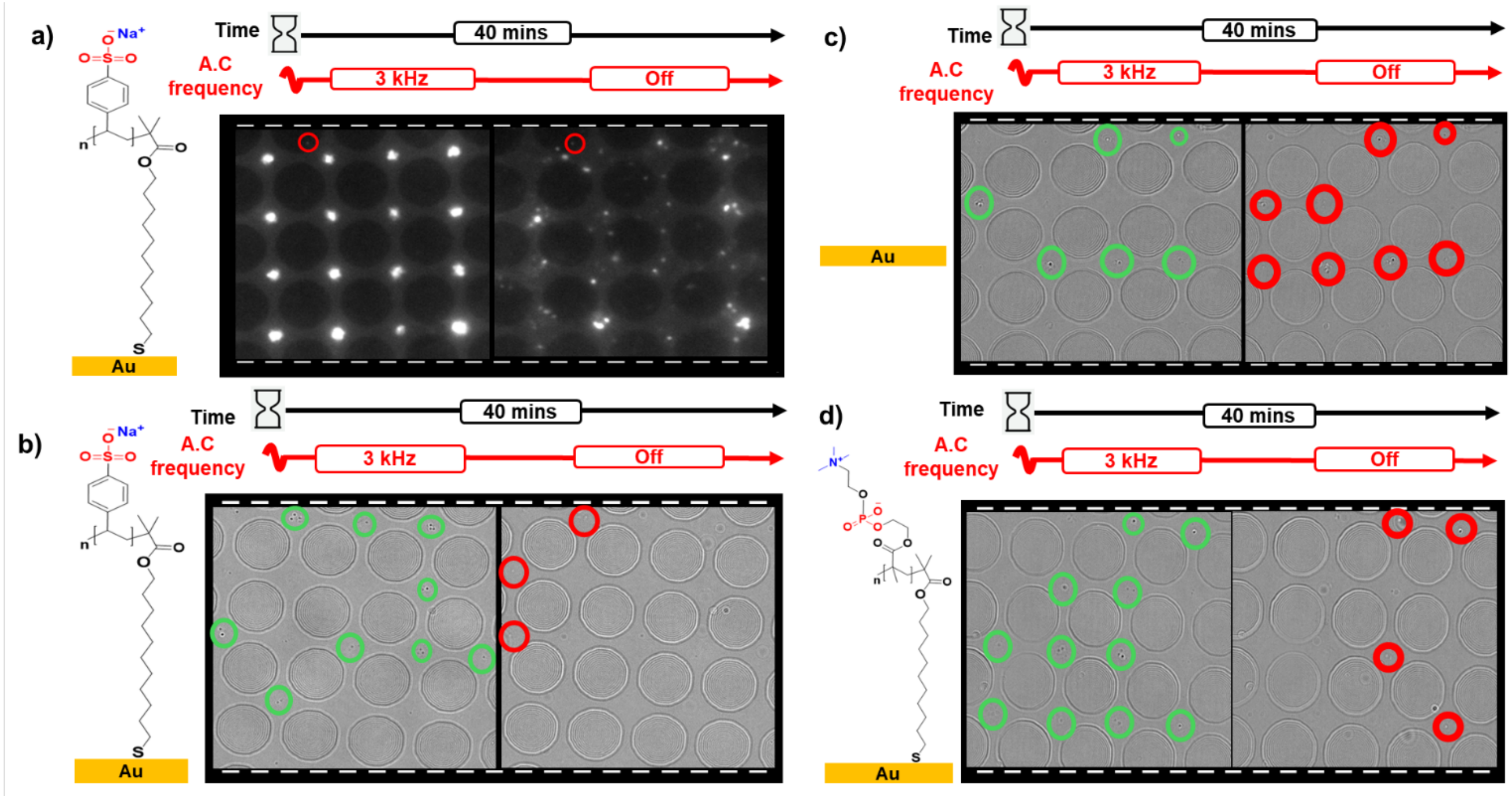

Figure 4: **a)** Fluorescence images of trapped and released polystyrene with IET functionalized with PSS **b)** ISCAT images of trapped and released (left and right) extracellular vesicles in a PSS functionalized IET **c)** unpassivated IET and **d)** IET passivated with PMPC. The green circles highlights some of the trapped particles at 3 kHz while the red circles highlight the adsorbed particles due to fouling.

Overrall, poly(sodium styrene sulphate) (PSS) with ATRP showed a significant reduction in fouling over MUA and unpassivated surface for polystyrene and have similar performance with zwitterionic PMPC for extracellular vesicles. Using ATRP technique as a method of surface passivation for nanotweezers not only enables that particles can be reversibly trapped and released at will, but also gives the leverage to vary the degree of hydrophilicity and surface charge by growing different types of antifouling monomers (terminal groups) for different particles of interest, extending the application of nanotweezers to a broader area.

Effective surface passivation requires consideration of several factors, including the operating pH of the antifouling material, the composition of the medium, and the solution conductivity. For example, mercaptoundecanoic acid (MUA) is most effective at pH values above 7, where its terminal carboxyl groups are deprotonated and negatively charged. The composition of the medium is also critical. In homogeneous media, such as those investigated in this work, knowledge of the particle zeta potential is generally sufficient to select an antifouling coating that electrostatically repels the particles. However, in heterogeneous media or when particle surface charges are unknown, antifouling materials with neutral terminal groups, such as zwitterionic compounds, are often preferred, as their performance is primarily governed by the formation of a strongly bound hydration layer that resists particle adhesion. Using ATRP protocol provides a versatile platform for tailoring these antifouling coatings by enabling precise control over polymer composition and film thickness, thereby allowing optimization for specific samples and experimental conditions. Finally, solution conductivity can significantly influence passivation performance. In systems that rely on hydration layer formation, high ionic strength can compress the electrical double layer at the surface [37], potentially compromising the stability and effectiveness of the passivated interface. Therefore, careful consideration of these factors is essential for achieving robust and reliable surface passivation in nanotweezer experiments.

## Conclusion

We successfully passivated the gold-based interferometric electrohydrodynamic tweezers (IET) with thiol-based antifouling materials using Atom Transfer Radical Polymerization (ATRP) for polystyrene and extracellular vesicle fouling reduction. Comparative analysis reveals that multiple layer PSS grown on a thiol initiator exhibits superior anti-fouling performance against polystyrene adsorption and gives a similar result with PMPC zwitterionic compound for extracellular vesicle antifouling. Our results suggests that MUA and PSS could serve as effective materials for reducing fouling of other materials with negative zeta potential such as nanoplastics. We have also shown that the ATRP method can be used to passivate a nanotweezer device with terminal groups tailored for different applications. With the appropriate surface initiators, this approach can be readily extended to a wide range of nanotweezer and biosensing substrates, including silicon, indium tin oxide, and silicon dioxide.

## Acknowledgments

The research reported in this publication was supported by the National Institute of General Medical Sciences of the National Institutes of Health under award number R35GM150572. We gratefully acknowledge Dr. Cal Craven and Soren Smail for his invaluable insights and contributions to this project.

## Supporting Information

- Movie 1: Milk EV trapped with bare IET after 40 minutes.
- Movie 2: Milk EV trapped with PMPC passivated IET after 40 minutes.
- Movie 3: 100 nm Polystyrene trapped with PSS passivated IET after 40 minutes.
- Movie 4: Milk EV trapped with PSS passivated IET after 40 minutes.
- Movie 5: 100 nm Polystyrene trapped with TMOL passivated IET after 40 minutes.
- Movie 6: 100 nm Polystyrene trapped with OEG passivated IET after 15 minutes.

## EXPERIMENTAL METHOD

### Materials

11-Mercaptoundecyl 2-Bromo-2-methylpropanoate, Sodium 4-vinylbenzenesulfonate (SS), 2-Methacryloyloxyethyl phosphorylcholine (MPC), 11-Mercaptoundecanoic acid (MUA), (1-Mercaptoundec-11-yl) tetra (ethylene glycol) (OEG), and 1-Mercapto-11-hydroxy-3,6,9-trioxaundecane (TMOL). These chemicals were obtained from sigma Aldrich and TCI chemicals and were used without further purification. A 100 nm fluorescence tagged polystyrene was purchased from Thermos Fisher Scientific.

### Extracellular vesicle Isolation

Commercially available 2% grade A reduced-fat bovine milk was obtained from a local grocery store. To avoid pH changes or protein degradation, all samples were processed within four hours of purchase. Raw milk was first centrifuged at 2,000 × g for 30 minutes at 4 °C to remove cream and coarse particulates. The clarified supernatant was carefully transferred into 50 mL conical tubes, ensuring the pellet fraction was not disturbed. To further eliminate residual fat droplets and casein aggregates, the supernatant was passed through a Whatman Grade 1 filter (150 mm, Cytiva, 18123982) prior to downstream ultracentrifugation.

To enrich for the whey fraction, casein was precipitated under acidic conditions. The defatted milk was pre-warmed at 37 °C for 10 minutes and then treated with 6 N acetic acid (Sigma-Aldrich, 102762795) at a ratio of 1:100 (acid:milk, v/v). Following 5 minutes of incubation at room temperature, the pH was confirmed at ~4.6. Samples were centrifuged at 10,000 × g for 10 minutes at 4 °C, after which the whey supernatant was collected, and the casein pellet discarded. The whey fraction was subsequently filtered through a 0.22 μm membrane filter (Millipore, 0000430715) to remove residual debris.

Extracellular vesicles (EVs) were recovered from the whey fraction by sequential ultracentrifugation. Samples were spun at 12,000 × g, 35,000 × g, 70,000 × g, and 100,000 × g for 90 minutes each at 4 °C using a Ti45 rotor (Optima XPN-80, Beckman Coulter). Pellets obtained from the 12K and 35K spins were discarded, while the 70K and 100K pellets were retained. The 100K pellet, enriched in EVs, was resuspended in phosphate-buffered saline (PBS, pH 7.4) and subjected to an additional washing step at 100,000 × g for 60 minutes at 4 °C in a TLA-110 rotor (Optima MAX-XP, Beckman Coulter). The final pellet was resuspended in 500 µL PBS. To further reduce protein aggregates, the EV suspension was clarified by centrifugation at 10,000 × g for 5 minutes at 4 °C, and the supernatant was collected as purified milk-derived EVs (mEVs). Aliquots were frozen at -20 °C in 1.5 mL microcentrifuge tubes (VMR, 87003-294) for further use.

**Sample Fabrication**

The 15-um hole array was fabricated using photolithography in the cleanroom. NR9 Photoresist was spin coated on an $O_2$ plasma-cleaned glass slide at 7000 rpm. The micro-hole pattern was then created on the photoresist using a Karl Suss MA-6 mask aligner and developed with MF13. Following the development, a descum step was performed to clean the unexposed photoresist residual before metal deposition. A 5 nm thick adhesion chromium and 100 nm gold film (15-nm Au for Interferometric imaging) were deposited on the developed samples using an E-beam deposition chamber (Angstrom Amod – Combined e-beam, Resistive & Sputter Deposition chamber). Lastly, NR9 was removed using DMSO at 80 $^0$C for 1 hour.

**Surface preparation: 11-Mercaptoundecanoic acid (MUA)**

The surface passivation technique follows a previously reported method [33] with certain modifications. The fabricated chips were cleaned with absolute ethanol and deionized water and then dried with a stream of $N_2$. The removal of organic contaminants was done by using a UV/Ozone cleaner for 30 minutes. The surface was rinsed again with absolute ethanol and deionized (DI) water and then dried with $N_2$. The cleaned chip was soaked in a 10 mM solution of the antifouling material for 24 hours. Finally, the chip was rinsed in ethanol and DI water and finally dried with $N_2$. The chip was then used to set up a microfluidic device for the trapping experiment.

**Surface preparation: Atom Transfer Radical Polymerization of PSS and PMPC**

We adapted the previously reported method [25] with some modification. The fabricated chips were cleaned with absolute ethanol and deionized water and then dried with a stream of $N_2$. The removal of organic contaminants was done by using UV/Ozone cleaner for 30 minutes. The surface was rinsed again with absolute ethanol and deionized (DI) water and then dried with $N_2$. The cleaned chip was soaked in a 10 mM solution of 11-Mercaptoundecyl 2-Bromo-2-methylpropanoate for 24 hours. The chip was rinsed with Tetrahydrofuran, ethanol and deionized water and then dried with $N_2$. The chip was placed in an Erlenmeyer flask containing Cu(I)Br, and $O_2$ gas was displaced with $N_2$ for 30 minutes. Separately, a 50/50 mixture of water and methanol solution of 2,2′-bipyridyl with either 0.1M sodium 4-vinylbenzenesulfonate (SS) or 0.1M 2-methacryloyloxyethyl phosphorylcholine (MPC) was simultaneously sparged with $N_2$. The two solutions were then combined and allowed to react for 24 hours under nitrogen protection. Finally, the chip was rinsed sequentially with ethanol, 1× PBS, and deionized water, and dried with $N_2$.

**Contact angle measurement.**

The contact angle of water on bare and functionalized IET with different antifouling materials were measured with Ossila contact angle goniometer at ambient temperature.

**ATR-FTIR Surface characterization**

Attenuated total reflectance Fourier transform infrared spectroscopy (ATR-FTIR) was performed using a Thermo Nicolet 6700 FT-IR spectrometer, equipped with a liquid nitrogen-cooled mercury–cadmium–telluride (MCT) detector and a Smart iTR™ ATR accessory with a diamond crystal plate. Functionalized spectra on IET surfaces were acquired with 256 scans collected at a resolution of 2 $cm^{-1}$.

**PM-IRRAS Surface characterization**

Using a Bruker PMA-50 adapter on a Bruker Tensor 27 infrared spectrometer, data for Polarization Modulation Infrared Reflection Absorption Spectroscopy (PM-IRRAS) were collected. The setup included a lock-in amplifier, a PEM-90 photoelastic modulator, and a mercury-cadmium-telluride (MCT) detector. Functionalized spectra on gold surfaces were collected at a resolution of 2 $cm^{-1}$ over a 20-minutes duration (see result in Supplementary Figure 1c).

**Particle concentration**

The particle concentrations, size and zeta potential were measured using Malvern Zetasizer.

**Fluorescence and ISCAT Microscopy**

Fluorescence and interferometric scattering (ISCAT) imaging experiments were used to compare the fouling reduction of polystyrene and extracellular vesicle in the IET system. All experiments were imaged using an inverted Nikon Ti2 microscope, and the acquired images were processed with ImageJ. The setup is shown in Supplementary Figure 1.